\documentstyle[aps,preprint,epsfig]{revtex}
\tightenlines
\newcommand{\bea}{\begin{eqnarray}}
\newcommand{\eea}{\end{eqnarray}}
\newcommand{\beq}{\begin{equation}}
\newcommand{\eeq}{\end{equation}}
\newcommand{\bay}{\begin{array}}
\newcommand{\eay}{\end{array}}
\begin{document}
\preprint{\parbox{6cm}{\flushright CLNS 98/1591\\TECHNION-PH-98-92\\[1cm]}}
\title{Weak phases $\gamma$ and $\alpha$ from $B^+$, or $B^0$ and $B_s$ decays}
\author{Michael Gronau}
\address{Physics Department,
Technion - Israel Institute of Technology, 32000 Haifa, Israel}
\author{Dan Pirjol}
\address{Floyd R. Newman Laboratory of Nuclear
Studies, Cornell University, Ithaca, New York 14853}
\date{\today}
\maketitle

\begin{abstract} 
An improved flavor SU(3) method is presented for determining the weak angle 
$\gamma$ of the unitarity triangle using decay rates for $B^+\to K\pi, 
B^+\to K^+\bar K^0$ and $B^+\to \pi^+\eta$ (or $B^0\to K\pi$ and $B_s\to K 
\pi$), their CP-conjugate modes and the CP-averaged rate for 
$B^\pm\to\pi^\pm\pi^{0}$. Rescattering (color-suppressed) contribution in 
$B^+(B^0)\to K\pi$, for which an improved bound is obtained, is subtracted 
away. The only significant SU(3) breaking effects are accounted for in the 
factorization approximation of tree amplitudes. The weak angle $\alpha$ is 
obtained as a byproduct. 
\end{abstract}
  
\pacs{pacs1,pacs2,pacs3}

\narrowtext

The determination of the angles of the unitarity triangle is an important goal 
of physics studies at existing and future $B$ meson facilities. It is 
expected to provide tests of the CKM mechanism of CP violation in the Standard 
Model and to shed light on possible new physics. In particular, the determination 
of the weak phase $\gamma=$Arg$(V_{ub}^*)$ has stimulated a great deal of 
effort, both on the theoretical and experimental side. A variety of ways have 
been proposed to extract this angle \cite{Rev}, ranging from theoretically 
clean methods \cite{BDK} applied to $D K$ modes \cite{Abi} hampered by some 
very small branching ratios, to approximate methods applied to $K \pi$ modes 
most of which have already 
been observed \cite{CLEO}. In the latter case one usually uses approximate 
flavor SU(3) symmetry of strong interactions \cite{GHLR} to relate $B \to 
K \pi$ to $B \to \pi \pi$ amplitudes. 

In a somewhat simplified version of this idea, Gronau, Rosner and London 
\cite{GRL} suggested to determine $\gamma$ through a triangle construction for 
$B^+$ decay amplitudes
into $K^0\pi^+, K^+\pi^0$ and $\pi^+\pi^0$, and for the corresponding 
charge-conjugate decays. It was later noted \cite{EWP} that higher order 
electroweak penguin (EWP) contributions upset this triangle construction. 
Various attempts were made to eliminate the uncertainties due to EWP amplitudes 
\cite{Rev}. Recently Neubert and Rosner \cite{NR2} included the EWP amplitudes 
in $B^+\to K\pi$ in a model-independent manner, by relating them to 
corresponding current-current amplitudes. (One often refers to such 
amplitudes as "tree" amplitudes, since they are of lowest order in electroweak 
couplings.) In their revised triangle construction the authors of \cite{NR2} must rely, 
however, on the dynamical assumption that the amplitude for $B^+\to K^0\pi^+$ 
is dominated completely by a QCD penguin contribution, and involves no term 
proportional to $e^{i\gamma}$ \cite{NR1}.
This assumption is equivalent to 
neglecting certain final state  rescattering effects \cite{Rescat}. Present 
experimental limits on such effects from SU(3) related $B\to K \bar K$ decays 
\cite{Falk,Limits} are not yet sufficiently strong for ignoring them. Indirect 
evidence against such effects could also be obtained from future limits on the 
CP asymmetry in $B^{\pm}\to K\pi^{\pm}$.

In view of the possibility that rescattering effects could give rise to a 
small, however non-negligible, contribution in $B^+\to K^0\pi^+$ with phase 
$\gamma$, thus upseting the construction of Ref.~\cite{NR2}, we propose in the 
present Letter to combine the processes $B^{\pm}\to K\pi$ with future 
information from $B^{\pm}\to K^{\pm} K$ and $B^{\pm}\to \pi^{\pm}\eta_8$ 
decays ($\eta_8$ is an SU(3) octet).
Alternatively, to avoid the question of $\eta-\eta'$ mixing, 
the same procedure can be applied by combining $B^0 \to K\pi$ and $B_s 
\to \bar K\pi$ decays. Using a simple SU(3) relation between 
these pairs of processes, we will show that one can avoid uncertainties 
due to final state rescattering in $B^+\to K^0\pi^+$ and due to a
color-suppressed amplitude in $B^0\to K^0\pi^0$.
SU(3) breaking effects occuring in these 
relations will be shown to contribute only a very small uncertainty in 
$\gamma$. SU(3) breaking, in the relation between tree amplitudes of $B\to 
K\pi$ and $B\to\pi\pi$, will be accounted for in the
factorization approximation. Electroweak penguin effects will be included in 
a model-independent way \cite{GPY}.
 
Using the notations of \cite{GPY}, we write the neutral and charged $B$ 
decay amplitudes into $K\pi$ states in terms of graphical SU(3) 
amplitudes
\bea
\label{2}
A(B^0\to K^+\pi^-) &=& |\lambda_u^{(s)}|e^{i\gamma} (-T-P_{uc}) -
                       |\lambda_t^{(s)}| (-P_{ct} + P_1^{EW})~,\\
\label{1}
\sqrt2 A(B^0\to K^0\pi^0) &=& |\lambda_u^{(s)}|e^{i\gamma} (-C+P_{uc}) -
                       |\lambda_t^{(s)}| (P_{ct} + \sqrt2 P_2^{EW})~,
\eea
\bea\label{B'1}
A(B^+\to K^0\pi^+) &=& |\lambda_u^{(s)}|e^{i\gamma} (A+P_{uc}) -
                       |\lambda_t^{(s)}| (P_{ct} + P_3^{EW})~,\\
\label{B'2}
\sqrt2 A(B^+\to K^+\pi^0) &=& |\lambda_u^{(s)}|e^{i\gamma} (-T-C-A-P_{uc}) -
                       |\lambda_t^{(s)}| (-P_{ct} + \sqrt2 P_4^{EW})~,
\eea
where $\lambda_{q'}^{(q)}=V^*_{q'b}V_{q'q}$, the amplitudes $T, C, A, P$ include
unknown strong phases, and $P_{1-4}^{EW}$ are the 
respective EWP contributions to these decays. The amplitudes 
(\ref{2})-(\ref{B'2}) satisfy the two triangle relations 
\cite{NR2,NQ}
\bea\nonumber
& &\sqrt2 A(B^+\to K^+\pi^0) + A(B^+\to K^0\pi^+) =
\sqrt2 A(B^0\to K^0\pi^0) + A(B^0\to K^+\pi^-) =\\
\label{3}
& &\qquad 
\sqrt2\lambda |A(B^+\to\pi^+\pi^0)|e^{i(\gamma+\phi)} \rho
\left(1 - \delta_{EW} e^{-i\gamma}\right)\,.
\eea
Here we denote $\lambda=V_{us}/V_{ud},~\delta_{EW}=
-(3/2)|\lambda^{(s)}_t/\lambda^{(s)}_u|\kappa \simeq 0.66$ ($\kappa\equiv
(c_9+c_{10})/(c_1+c_2)=-8.8\cdot 10^{-3}$), while $\phi$ is an unknown strong phase. 
The second term in the brackets represents the sum of EWP contributions to the 
amplitudes on the left-hand-sides \cite{NR2,GPY}.
The correction factor $\rho=(f_K/f_\pi) |1+(3/2)\kappa|
\lambda_t^{(d)}/\lambda_u^{(d)}|\exp(i\alpha)|^{-1}\approx 1.22$ accounts for 
factorizable SU(3) breaking effects and EWP contributions
to the amplitude $A(B^+\to\pi^+\pi^0)$ respectively. Numerically this factor is 
dominated by the former contribution.

Each of the two amplitude triangles (\ref{3}) cannot be used by 
itself, together with the corresponding relation for the CP-conjugate 
amplitudes $\tilde A(\bar B\to\bar f) \equiv e^{2i\gamma} A(\bar B\to\bar f)$, 
to allow a determination of $\gamma$. The reason is that in general all four 
amplitudes in (\ref{2})-(\ref{B'2}) involve two terms with different weak 
phases. ~$\gamma$ can be determined only when one of these terms can be 
neglected in one of the amplitudes as assumed in \cite{GRL,NR2}. Although the 
first terms in (\ref{1}) and (\ref{B'1}) are likely to be significantly 
smaller than 
the second terms, we will not neglect them in the forthcoming discussion. For 
definiteness, we present in detail the version of our method applied to 
neutral $B$ decays. A similar brief treatment of $B^+$ decays precedes
the conclusion.  

Normalizing amplitudes by $A(B^+\to\pi^+\pi^0)$, we define reduced amplitudes 
\bea
x_{+-} &=& \frac{1}{\sqrt2\lambda\rho}\frac{|A(B^0\to K^+\pi^-)|}
{|A(B^+\to\pi^+\pi^0)|}~,\qquad
x_{00} = \frac{1}{\lambda\rho}\frac{|A(B^0\to K^0\pi^0)|}
{|A(B^+\to\pi^+\pi^0)|}~,\\
\tilde x_{-+} &=& \frac{1}{\sqrt2\lambda\rho}
\frac{|A(\bar B^0\to K^-\pi^+)|}{|A(B^+\to\pi^+\pi^0)|}~,\qquad
\tilde x_{00} = \frac{1}{\lambda\rho}
\frac{|A(\bar B^0\to \bar K^0\pi^0)|}{|A(B^+\to\pi^+\pi^0)|}~.
\eea
The triangle relation (\ref{3}) for $B^0$ decays and its $CP$-conjugate are 
given by
\bea\label{tr1}
& &x_{00}e^{i\phi_1} + x_{+-}e^{i\phi'_1} = 1 - \delta_{EW} e^{-i\gamma}~,\\
\label{tr2}
& &\tilde x_{00}e^{i\tilde\phi_1} + \tilde x_{-+}e^{i\tilde\phi'_1} = 1 - 
\delta_{EW} e^{i\gamma}~,
\eea
where $\phi_1, \phi'_1, \tilde\phi_1, \tilde\phi'_1$ contain both strong and 
weak phases. These triangles are represented in Fig.~1. 
The angles $\phi_1$ and $\tilde\phi_1$ are functions of $\cos\gamma$ defined by  
second order equations
\bea\label{phi1}
& &(1-\delta_{EW}\cos\gamma)\cos\phi_1 + \delta_{EW}\sin\gamma\sin\phi_1 =
\frac{x^2_{00}-x^2_{+-}+(1+\delta_{EW}^2-2\delta_{EW}\cos\gamma)}{2x_{00}}~,\\
\label{phi2}
& &(1-\delta_{EW}\cos\gamma)\cos\tilde\phi_1 - \delta_{EW}\sin\gamma\sin\tilde
\phi_1 = \frac{\tilde x^2_{00}-\tilde x^2_{-+}+(1+\delta_{EW}^2-2\delta_{EW}
\cos\gamma)}{2\tilde x_{00}}~.
\eea

As mentioned, a major simplification occurs when the color-suppressed 
amplitude $-C+P_{uc}$ in $A(B^0\to K^0\pi^0)$ (\ref{1}) is neglected relative 
to the dominant penguin contribution.
In this limit, the angle between the amplitudes $\sqrt2 A(B^0\to K^0\pi^0)$ 
and $\sqrt2 \tilde A(\bar B^0\to \bar K^0\pi^0)$ in Fig.~1 is $2\gamma$. This 
implies $\cos 2\gamma = \cos(\tilde\phi_1-\phi_1)$ which determines $\gamma$. 
A similar argument applies in charged $B$ decays \cite{NR2} when
the annihilation amplitude $A+P_{uc}$ is neglected in $A(B^+\to K^0\pi^+)$
(\ref{B'1}). In general, without neglecting these terms, the two triangles in 
Fig.~1 involve an arbitrary relative angle which prohibits a determination of
$\gamma$.

In order to avoid these dynamical assumptions and to establish another
constraint on the relative angles between the above two triangles, let us 
consider together with $B^0\to K\pi$  also the following $B_s$ decay amplitudes 
\cite{GPY}
\bea\label{5}
A(B_s\to K^-\pi^+) &=& |\lambda_u^{(d)}|e^{i\gamma} (-T'-P'_{uc}) +
                       |\lambda_t^{(d)}|e^{-i\beta} (-P'_{ct} + P_1^{'EW})~,\\
\label{4}
\sqrt2 A(B_s\to \bar K^0\pi^0) &=& |\lambda_u^{(d)}|e^{i\gamma} (-C'+P'_{uc}) +
                     |\lambda_t^{(d)}|e^{-i\beta} (P'_{ct} + \sqrt2 P_2^{'EW})~.
\eea
In the SU(3) symmetric limit the reduced amplitudes appearing in these 
expressions are
equal to those appearing in Eqs.~(\ref{2}) and (\ref{1}), $T=T'\,, 
C=C'\,, P_{uc}=P'_{uc}\,,P_{ct}=P'_{ct}\,, P_1^{EW}=P_1^{'EW}\,,P_2^{EW}=
P_2^{'EW}$. (We will consider below uncertainties due to this approximation).
In the first case this follows simply from U-spin. The amplitudes 
(\ref{5}) and (\ref{4}) satisfy a triangle relation similar to (\ref{3})
\bea\label{6}
\sqrt2 A(B_s\to \bar K^0\pi^0) + A(B_s\to K^-\pi^+) = \sqrt2 \rho' 
A(B^+\to\pi^+\pi^0)~.
\eea
This relation is exact, even accounting for EWP contributions. The factor 
$\rho'=((M_{B_s}^2-M_K^2)F_{B_sK}(M_K^2))/((M_{B}^2-M_\pi^2)F_{B\pi}
(M_\pi^2))$ parametrizes the leading factorizable SU(3) breaking effects.

The SU(3) relations between the terms of definite CKM factors in 
(\ref{2}), (\ref{1}) and in (\ref{5}),(\ref{4}), respectively, allow a simple 
geometrical interpretation. Drawing the amplitudes (\ref{1}), (\ref{4})-scaled 
by $\lambda$ and their CP-conjugates ($\tilde A(\bar B\to\bar f) 
\equiv e^{2i\gamma} A(\bar B\to\bar f)$), such that all amplitudes originate in 
a common point, the other ends of the four amplitudes form a quadrangle as 
shown in Fig.~2. (The point of origin is not shown in this figure. In Fig.~1 it 
is chosen as the point $O$.) This 
quadrangle is not determined by rate measurements alone, since it involves the 
unknown relative angle between the triangles (\ref{3}), (\ref{6}) and their 
charge-conjugates, which depends on $\gamma$ through $\phi_1, \tilde\phi_1$. 
We will show now that the quadrangle provides another condition on $\gamma$ 
which fixes this phase.

Consider the four sides of the quadrangle in Fig.~2 given in the SU(3) limit 
by (with $p\equiv P_{ct}+\sqrt2 P_2^{EW}$)
\bea\label{Quadr}\nonumber
v &=& |\lambda_t^{(s)}|(1-e^{2i\gamma})p~,\qquad\qquad\qquad\,\,
x = \left(\lambda|\lambda_t^{(d)}|e^{i(\beta+2\gamma)}+|\lambda_t^{(s)}|\right)
p~,\\
z &=& \left(\lambda|\lambda_t^{(d)}|e^{-i\beta}+|\lambda_t^{(s)}|e^{2i\gamma}
\right)p~,\qquad
y = \lambda|\lambda_t^{(d)}|\left(e^{i(\beta+2\gamma)}-e^{-i\beta}\right)p~.
\eea
Since all four sides of the quadrangle are proportional to a {\em single} 
hadronic amplitude $p\equiv P_{ct}+\sqrt2 P_2^{EW}$, its shape is determined 
exclusively by CKM parameters. In fact this quadrangle is an 
isosceles trapezoid, $|x|=|z|$, whose sides $v$ and $y$ are parallel. We will 
select a point $X$ on the median of the trapezoid (the line
bisecting the sides $v$ and $y$ perpendicularly) with the property that
its distances to the vertices of the trapezoid  are in the following 
ratio
\bea\label{r}
r = \frac{AX}{CX} = \frac{BX}{DX} = \frac{|\lambda_t^{(s)}|}{\lambda|
\lambda_t^{(d)}|} = \frac{1}{\lambda}\frac{|V_{ts}|}{|V_{td}|}=22 \pm 4~.
\eea
The value of $|V_{ts}/V_{td}|$ is taken from a recent global analysis of
the unitarity triangle \cite{Rosner}.
It is easy to see that the angles through which the sides $v$ and $y$ are 
seen from the point $X$ are $2\gamma$ and  $2\alpha$, respectively. (See 
Fig.~2).  The 
new condition on $\gamma$, together with 
(\ref{phi1}), (\ref{phi2}) illustrated in Fig.~1, are sufficient for determining 
this phase up to discrete ambiguities (to be discussed below). Fig.~2 can also 
be used to measure $\alpha$.

The conditions (\ref{r}) can be applied to determine the point $X$ in Fig.~1
in the following way.  First, we note that the points $C$ and $D$ are fixed by
Eq.~(\ref{6}) and its charge-conjugate. Then, recall that the set of points 
$X$, for which the
ratio of the distances to two given points $A$ and $C$ takes a fixed value 
$r$, is a circle given by
\bea\label{circle}
\left|X - \frac{Cr^2-A}{r^2-1}\right| = \frac{r|A-C|}{r^2-1}~.
\eea
Using $r^2\gg 1,~|A|\sim 20|C|$ and $|C|r^2\gg |A|$ (see discussion below), 
where $A$ and $C$ are the coordinates of these points with respect to the 
origin $O$ shown in Fig.~1, the circle is approximated by 
$|X-C|=|A-C|/r \simeq |A|/r$.
The second condition (\ref{r}), applied to $B$ and $D$, has a similar form, 
$|X-D|=|B-D|/r \simeq |B|/r$. The two circles of equal radii, $|A|/r
\approx |B|/r$, with centers at $C$ and $D$, intersect at $X$ and determine this 
point up to a possible two-fold ambiguity. $\gamma$ is generally determined 
up to an eight-fold ambiguity due to an additional up-down ambiguity 
of the two $B_s$ triangles. In practice, four of these 
possibilities might be eliminated if the two circles do not intersect.

In order to demonstrate the algebraic solution for the two weak phases, let us 
introduce also reduced amplitudes for $B_s$ decays 
\bea
y_{-+} &=& \frac{1}{\sqrt2\rho'}\frac{|A(B_s\to K^-\pi^+)|}
{|A(B^+\to\pi^+\pi^0)|}~,\qquad
y_{00} = \frac{1}{\rho'}\frac{|A(B_s\to\bar K^0\pi^0)|}{|A(B^+\to\pi^+\pi^0)|}~,
\\
\tilde y_{+-} &=& \frac{1}{\sqrt2\rho'}\frac{|A(\bar B_s\to K^+\pi^-)|}
{|A(B^+\to\pi^+\pi^0)|}~,\qquad
\tilde y_{00} = \frac{1}{\rho'}\frac{|A(\bar B_s\to K^0\pi^0)|}{|A(B^+\to
\pi^+\pi^0)|}~.
\eea
These amplitudes satisfy the triangle relations
\bea\label{tr3}
y_{00}e^{i\phi_2} + y_{-+}e^{i\phi'_2} = 1~,\qquad
\tilde y_{00}e^{i\tilde\phi_2} + \tilde y_{+-}e^{i\tilde\phi'_2} = 1~,
\eea
where tiny EWP contributions to the amplitude $A(B^+\to\pi^+\pi^0)$ are 
neglected. (Their effects will be estimated below). 
The phases $\phi_2$ and $\tilde\phi_2$ are determined from rate measurements 
through (\ref{tr3})
\bea\label{phi3}
\cos\phi_2 = \frac{1+y_{00}^2-y_{-+}^2}{2y_{00}}~,\qquad
\cos\tilde\phi_2 = \frac{1+\tilde y_{00}^2-\tilde y_{+-}^2}{2\tilde y_{00}}~.
\eea

The angle $\gamma$ is extracted as the root of the equation $\cos(BXA) = 
\cos2\gamma$. Denoting the position of the point $X$ by $\rho e^{i\phi}$, 
determined as explained above, an explicit form for this equation is
\bea\label{eq1}
2x_{00}\tilde x_{00}\sin(\tilde\phi_1-\phi_1) \sin 2\gamma -
2[\rho^2-x_{00}\tilde x_{00}\cos(\tilde\phi_1-\phi_1)]\cos 2\gamma =
x_{00}^2 + \tilde x_{00}^2 - 2\rho^2~.
\eea
This determines $\gamma$ when combined with Eqs.~(\ref{phi1})(\ref{phi2}).
The angle $\alpha$ is given directly by the angle $CXD$, 
\bea
\sin\alpha = \frac12\frac{|y_{00} e^{i\phi_2} - \tilde y_{00} e^{i\tilde \phi_2}|}
{x_{00}/r}~.
\eea

In order to evaluate the precision of this method, let us first consider the  
magnitudes of the amplitude ratios appearing in the triangle relations 
(\ref{tr1}), (\ref{tr2}) and (\ref{tr3}). Using the measured decay rates
of $B\to K\pi$ and $B\to \pi\pi$ \cite{CLEO}, one estimates from the dominant 
terms \cite{NR1,Limits} $\tilde x_{00}\simeq x_{00}\simeq\tilde 
x_{-+}\simeq x_{+-}
\simeq |\lambda_t^{(s)}P_{ct}/\lambda_u^{(s)}(T+C)|\simeq 4$. Similarly,
$\tilde y_{+-}\simeq y_{-+}\simeq 1,~\tilde y_{00}\simeq y_{00}\simeq |C/T|
\simeq 0.2$. We also note that since $BX/CX=|\lambda_t^{(s)}|/\lambda|
\lambda_t^{(d)}|=r$, the isosceles triangle in Fig. 2 with angle $2\gamma$ is 
about 20 times larger than the one with angle $2\alpha$.
These estimates justify the approximations made below 
Eq.~(\ref{circle}). The errors in the distances of the center points of the 
two circles from $O$, and the errors in the radii of these circles are
each of order $x_{00}/r^2\simeq 0.01$ and can be neglected. 

We now discuss the theoretical errors in the determination of $\gamma$ and 
$\alpha$. An intrinsic source of uncertainty is the parameter 
$\delta_{EW}\simeq 
0.63\pm 0.11$
\cite{NR2}, where the 5\% shift from 0.66 accounts for factorizable 
SU(3)-breaking corrections, and the error is dominated by the present 
poorly known ratio of CKM
matrix elements $|V_{ub}/V_{cb}|$. Its effect on the extraction of $\gamma$
was examined in detail in \cite{NR2}, and we have nothing new to add to
that discussion.

We will focus instead on the SU(3) breaking effects introduced by the additional
amplitudes considered in this method. They show up as differences between the 
amplitudes contributing to $B^0 \to K^0\pi^0$ (\ref{1}) and $B_s\to\bar K^0\pi^0$ 
(\ref{4}),
$|c'|\neq |c|$ (with $c\equiv C-P_{uc}$) and analogous inequalities holding 
for the corresponding 
penguin amplitudes, $|p'|\neq |p|$ (with $p\equiv P_{ct}+\sqrt2 P^{EW}_2$). 
One expects these amplitudes to differ by 
at most 30\%. A smaller uncertainty exists in the factor $\rho'$, for which 
the deviation from unity can be taken from quark models. Fixing $|p|$ 
and $|c|$ and allowing $|p'/\rho'|$ and $|c'/\rho'|$ to vary
within 40\%, the points $C$ and $D$ in Fig.~2 can vary within small circles of 
radius $\Delta y_{00}\simeq 0.4y_{00}\simeq 0.08$. Therefore, the corrections 
to $\gamma$ due to SU(3) breaking are expected to be small. (This is due to 
our judicious choice of origin about which the triangles in Fig.~1 are 
rotated, 
this point being adjacent to the color-suppressed amplitude for $B_s$ decay). 
To estimate the absolute value of the error in $\gamma$ arising from SU(3) 
breaking, we consider the most unfavorable case of a simultaneous shift of 
$y_{00}$ and $\tilde y_{00}$ by $\Delta y_{00}=0.08$. This 
translates into an error in $\gamma$ of $\Delta\gamma\simeq \Delta 
y_{00}/x_{00} \simeq 0.02$ which is about $1^{\circ}$.

The ratio $r$ introduced in (\ref{r}) is known with an error of about 20\%. 
This affects the determined position of the point $X$ through the radii of the
circles centered at $C$ and $D$.
The radii of these circles are of the order of $x_{00}/r\simeq 0.2$, implying
an error in the position of the point $X$ of the order of 0.04, which is half 
of the uncertainty arising from SU(3) breaking. 
Combining these two errors in quadrature, one obtains a total error in $\gamma$ 
of about $1.3^{\circ}$.

Another source of theoretical uncertainty is connected with the neglect of EWP 
contributions in $A(B^+\to\pi^+\pi^0)$. We have recently shown that when these 
effects are included, the relation between this decay amplitude and its CP 
conjugate is \cite{GPY,BuFl}
\bea
A(B^+\to \pi^+\pi^0) = e^{2i\xi} \tilde A(B^-\to\pi^-\pi^0)~,\qquad
\tan\xi = \frac{x\sin\alpha}{1+x\cos\alpha}~,
\eea
where $x=-(3/2)\kappa\sin\alpha/\sin(\alpha+\gamma)$. Numerically the angle 
$2\xi$ is seen to be very small, under $2^{\circ}$. This uncertainty will 
affect only the relative orientation of the two $B_s$ triangles, shifting the 
angles $\phi_2$ and $\tilde\phi_2$ by an amount $\Delta \phi_2= - 
\Delta\tilde\phi_2 = \xi$.
(The effect of these EWP on the $B\to K\pi$ triangles (\ref{3}) enters only 
through the 
factor $\rho$, to which they contribute at the level of 1\%).
The corresponding error in the positions of $C$ and $D$ is of order 
$0.2\xi\simeq 0.003$ which is well under the uncertainty arising from the 
other sources discussed above.

On the other hand, the smallness of the $CXD$ triangle implies that SU(3) 
breaking effects will have a larger impact on the extraction of $\alpha$ from 
this method. The estimates given above indicate that the error in such a 
determination is at the level of 30\%.

A similar method can be applied to the determination of $\gamma$ from
$B^+\to  K^0\pi^+$ and $B^+\to K^+\pi^0$ decays. In this case 
uncertainties due to rescattering in $B^+\to K^0\pi^+$ can be eliminated by 
considering in 
addition the decays $B^+\to K^+ \bar K^0$ and $B^+\to \pi^+\eta_8$,
where $\eta_8$ is an SU(3) octet. Their amplitudes are given by \cite{GHLR,GPY}
\bea\label{K+K}
A(B^+\to K^+\bar K^0) &=& |\lambda_u^{(d)}|e^{i\gamma} (A+P_{uc}) +
|\lambda_t^{(d)}|e^{-i\beta}(P_{ct} + P_3^{EW})~,\\
\sqrt{6}A(B^+\to \pi^+\eta_8) &=& |\lambda_u^{(d)}|e^{i\gamma} 
(-T-C-2A-2P_{uc}) +
|\lambda_t^{(d)}|e^{-i\beta}(-2P_{ct} + P_5^{EW})~,
\eea
and are closely related to (\ref{B'1}) and (\ref{B'2}). Their relative
orientation with respect to (\ref{B'1}) and (\ref{B'2}) can be fixed as in
the $B^0$ case with the help of the (exact) triangle relation
\bea
A(B^+\to K^+\bar K^0) + \sqrt{\frac32}A(B^+\to \pi^+\eta_8) =
\frac{1}{\sqrt2}A(B^+\to \pi^+\pi^0)~.
\eea
This triangle relation replaces Eq.~(\ref{6}) in the case of neutral $B$ 
decays. Instead of the quadrangle of Eqs.~(\ref{Quadr}), one now constructs a 
quadrangle from $A(B^+\to K^0\pi^+),~\lambda A(B^+\to K^+\bar K^0)$ and their 
charge-conjugates, the four sides of which are all proportional to $P_{ct} +
P_3^{EW}$. The extraction of $\gamma$ and $\alpha$ follows in a similar way. 
This set of processes is experimentally more accessible than $B_s\to\bar 
K^0\pi^0$, however $B^+\to \pi^+\eta_8$ involves a certain amount of 
model-dependence related to $\eta-\eta'$ mixing \cite{GReta}. 

One can use the arguments presented here, with figures similar to the above
drawn for $B^+$ decay amplitudes, to obtain an upper bound on rescattering 
effects in $B^+\to K^0\pi^+$ in terms of the charge-averaged 
$B^\pm\to K^\pm\bar K^0$ rate. The amplitudes of interest in $B^+$ decays are 
given by line segments analogous to those in Fig.~1.
(no normalization by $\sqrt2 A(B^+\to\pi^+\pi^0)$ is used.)
\beq
|OX| = |\lambda_u^{(s)}(A+P_{uc})|\,,\qquad
|XA| = |\lambda_t^{(s)}(P_{ct}+P_3^{EW})|~.
\eeq 
Simple geometry implies
\bea\nonumber
\epsilon_A &\equiv& \left|\frac{\lambda_u^{(s)}(A+P_{uc})}{\lambda_t^{(s)}(P_{ct}+P_3^{EW})}
\right| = \frac{|OX|}{|XA|}
\leq \frac{\mbox{min}(|OC|, |OD|) + \frac{1}{r}|OA|}{|OA|}
\leq \frac{1}{r} + 
\lambda\sqrt{\frac{B(B^\pm\to K^\pm \bar K^0)}{B(B^\pm\to K^0 \pi^\pm)}}~.\\
~\eea
The ratio $\epsilon_A$, describing rescattering in $B^+\to K^0\pi^+$, takes 
its maximum value when $|OC|=|OD|$ (for fixed $|OC|^2+|OD|^2$).
The expression on the right-hand side is accurate up to 
corrections of order $|OX|/|OA|\simeq 0.05$ of its magnitude (due to the
approximation $|XA| \simeq |OA|$ used in the second step). A previous bound
\cite{Falk}, based on the assumption of constructive interference between
the two terms in Eq.~(\ref{K+K}), omitted the $1/r$ term.

In conclusion, we have presented a new method for extracting the weak angle
$\gamma$ using combined $B^0$ and $B_s$ decays, or combining $B^+\to K\pi$
with $B^+\to K^+\bar K^0$ and $B^+\to \pi^+\eta$. This method represents an 
improvement of the method suggested in \cite{NR2} in that color-suppressed 
contributions in $B^0$ decay, or rescattering effects in case of $B^+$ decay, 
are eliminated with the help of SU(3) flavor symmetry. The additional
SU(3) breaking corrections were shown to be negligible.
Under ideal experimental conditions, this method would allow a substantial 
improvement in the precision of determining $\gamma$. In reality, $B_s$ 
decay modes involving neutral pions pose a particularly difficult experimental 
challenge. Alternatively, the use of charged $B$ decays involves
a slight theoretical complication due to $\eta-\eta'$ mixing which must be 
resolved.

{\em Acknowledgements.} D. P. would like to thank Peter Gaidarev, Tung-Mow 
Yan and Piljin Yi for useful discussions. We are grateful to Kaustubh Agashe
for pointing out an error in an early version of the paper.
This work is supported by the National Science 
Foundation and by the United States - Israel Binational Science Foundation 
under Research Grant Agreement 94-00253/3.

\newpage
\thispagestyle{plain}
\begin{figure}[hhh]
 \begin{center}
 \mbox{\epsfig{file=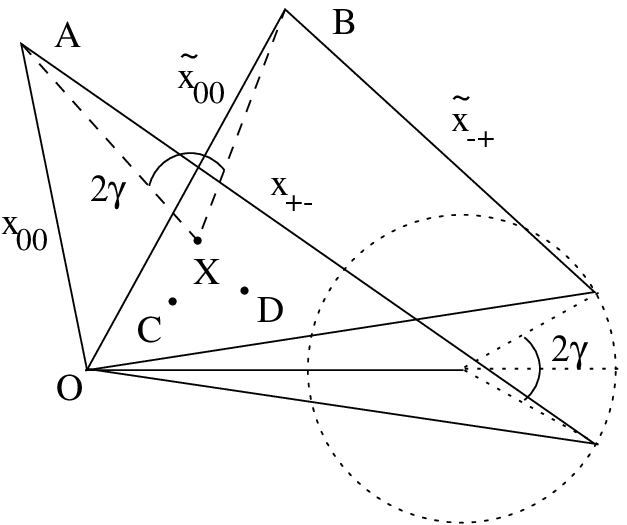,width=7cm}}
 \end{center}
 \caption{Relative orientation of $B^0$ amplitude triangles. $C$ and $D$ are 
the tips of the $B_s$ triangles (not shown for clarity), which 
determine the point $X$ as explained in the text below Eq.~(\ref{circle}) 
(see also Fig.~2).}
\label{fig1}
\end{figure}

\begin{figure}[hhh]
 \begin{center}
 \mbox{\epsfig{file=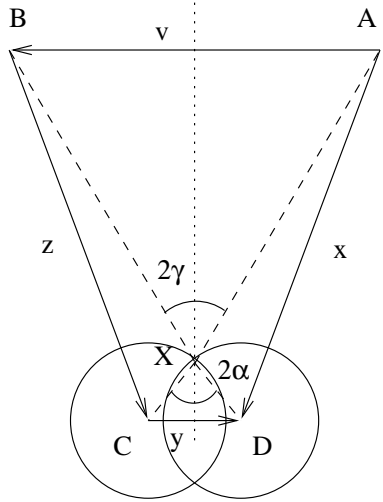,width=5cm}}
 \end{center}
 \caption{Quadrangle formed by the tips of the triangles for $B^0$ and $B_s$
decays $A=\sqrt2 A(B^0\to K^0\pi^0)$, $B=\sqrt2 \tilde A(\bar B^0\to\bar 
K^0\pi^0)$, $C=\lambda \sqrt2 A(B_s\to\bar K^0\pi^0)$ and $D=\lambda 
\sqrt2\tilde A(\bar B_s\to K^0 \pi^0)$. The point $X$ is determined by the
intersection of the two circles of radius $x_{00}/r\simeq \tilde x_{00}/r$ 
centered at $C$ and $D$ respectively.}
\label{fig2}
\end{figure}

\end{document}